\documentclass{article}
\usepackage{spconf,amsmath,graphicx,hyperref,xcolor,amsfonts}

\hypersetup{
    colorlinks=true,
    citecolor=blue,
    linkcolor=blue,
    filecolor=magenta,      
    urlcolor=blue,
}
\usepackage{enumitem}
\setlist[1]{itemsep=1pt}

\usepackage{booktabs}
\graphicspath{{figures/}} 

\title{Multi-dimensional Speech Quality Assessment in Crowdsourcing}
\name{Babak Naderi, Ross Cutler, Nicolae-C\u{a}t\u{a}lin Ristea}
\address{Microsoft Corporation, Redmond, USA \\
    babaknaderi $\vert$ ross.cutler $\vert$ nristea @microsoft.com}
\begin{document}
\maketitle

\begin{abstract}
Subjective speech quality assessment is the gold standard for evaluating speech enhancement processing and telecommunication systems. The commonly used standard ITU-T Rec.~P.800 defines how to measure speech quality in lab environments, and ITU-T Rec.~P.808 extended it for crowdsourcing. ITU-T Rec.~P.835 extends P.800 to measure the quality of speech in the presence of noise. ITU-T Rec.~P.804 targets the conversation test and introduces perceptual speech quality dimensions which are measured during the listening phase of the conversation. The perceptual dimensions are noisiness, coloration, discontinuity, and loudness. We create a crowdsourcing implementation of a multi-dimensional subjective test following the scales from P.804 and extend it to include reverberation, the speech signal, and overall quality. We show the tool is both accurate and reproducible. The tool has been used in the ICASSP 2023 Speech Signal Improvement challenge and we show the utility of these speech quality dimensions in this challenge. The tool will be publicly available as open-source at \url{https://github.com/microsoft/P.808}.
\end{abstract}

\vspace{2mm}
\noindent\textbf{Index Terms}: speech quality assessment, subjective test, crowdsourcing, perceptual dimensions, signal quality.

\section{Introduction}
\label{sec:intro}
Audio telecommunication systems, such as remote collaboration systems, smartphones, and telephones, are now ubiquitous and essential tools for work and personal use. Audio engineers and researchers have been working to improve the speech quality of these systems, with the goal of making them as good or better than face-to-face communication. However, there is still room for improvement, as it is still common to hear frequency response distortions, isolated and non-stationary distortions, loudness issues, reverberation, and background noise in audio calls.

Subjective speech quality assessment is the gold standard for evaluating speech enhancement processing and telecommunication systems, and the ITU-T has developed several recommendations for subjective speech quality assessment. ITU-T P.800 \cite{noauthor_itu-t_1996}  describes lab-based methods for the subjective determination of speech quality, including the Absolute Category Rating (ACR). ITU-T P.808 \cite{noauthor_itu-t_2018} describes a crowdsourcing approach for conducting subjective evaluations of speech quality. It provides guidance on test material, experimental design, and a procedure for conducting listening tests in the crowd. The methods are complementary to laboratory-based evaluations described in P.800. An open-source implementation of P.808 is described in \cite{naderi_open_2020}. ITU-T P.835 \cite{noauthor_itu-t_2003} provides a subjective evaluation framework that gives standalone quality scores of speech (SIG) and background noise (BAK) in addition to the overall quality (OVRL). An open-source implementation of P.835 is described in \cite{naderi_subjective_2021}. Perceptual dimensions for speech quality are identified in \cite{waltermann2006underlying} and extended to be noisiness, coloration, discontinuity and loudness~\cite{waltermann2013dimension}. Those are extensively studied in conversational test \cite{koster_identifying_2017, koster2018multidimensional,koster2015perceptual} and are focus of more recent multidimensional speech quality assessment standards namely ITU-T P.863.2 \cite{noauthor_itu-t_2022} and P.804 \cite{noauthor_itu-t_2017-1} (listening phase) (Table \ref{tab:evaluation}).

Intrusive objective speech quality assessment tools such as Perceptual Evaluation of Speech Quality (PESQ) \cite{rix_perceptual_2001} and Perceptual Objective Listening Quality Analysis (POLQA) \cite{beerends_perceptual_2013} require a clean reference of speech. Non-intrusive objective speech quality assessment tools like ITU-T P.563 \cite{noauthor_itu-t_563} do not require a reference, though it has low correlation to subjective quality \cite{avila_non-intrusive_2019}. Newer neural net-based methods, such as \cite{avila_non-intrusive_2019, reddy_dnsmos_2021, reddy_dnsmos_2022, yi_conferencingspeech_2022} provide better correlations to subjective quality. NISQA \cite{mittag_nisqa_2021} is an objective metric for P.804, but the correlation to subjective quality is not sufficient to use as a challenge metric.

Lab-based subjective testing in practice is slow due to the recruitment of test subjects and the limited number of test subjects, and expensive due to paying qualified test subjects and the cost of the test lab. The speed and cost result in the vast majority of research papers not using subjective tests but rather objective functions that are not well correlated to subjective opinion. An alternative to lab-based subjective tests is to crowdsource the testing. We introduce a crowdsourced multi-dimensional speech quality assessment tool that extends P.804 by adding SIG, OVRL, and reverberation (see Table \ref{tab:evaluation}). We show the tool is both accurate compared to lab results and is reproducible. The tool has been successfully used in the ICASSP 2023 Speech Signal Improvement challenge \cite{cutler_icassp_2023-1}.

In Section \ref{sec:implementation} we describe the implementation of the tool. In Section \ref{sec:validation} we provide accuracy and reproducibility analysis. In Section \ref{sec:usage} we provide an example usage of the tool. In Section \ref{sec:conclusions} we discuss conclusions and future work. 

\begin{table*}[h]
    \caption{Speech quality areas from P.804 listening phase (the first four) plus three additional areas}
    \label{tab:evaluation}
    \setlength\tabcolsep{2.0pt}
    \centering
    \begin{tabular}{c c c }
        \toprule
        Area & Description & Possible source \\
        \midrule
        Noisiness & Background noise, circuit noise, coding noise; BAK & Coding, circuit or background noise; device \\
        Coloration & Frequency response distortions & Bandwidth limitation, resonances, unbalanced freq. response \\
        Discontinuity & Isolated and non-stationary distortions & Packet loss; processing; non-linearities \\
        Loudness & Important for the overall quality and intelligibility & Automatic gain control; mic distance \\
        Reverberation & Room reverberation of speech and noise & Rooms with high reverberation \\
        Speech Signal & Overall signal quality & \\
        Overall & Overall quality& \\
     \bottomrule
    \end{tabular}
    \vspace{-0.2cm}
\end{table*}

\section{Implementation}
\label{sec:implementation}

We extended the P.808 Toolkit\cite{naderi_open_2020} to include a test template for a multi-dimensional quality assessment. The toolkit provides scripts for preparing the test, including packing the test clips in small test packages, preparing the reliability check questions, and analyzing the results. We ask participants to rate the perceptual quality dimensions of speech namely coloration, discontinuity, noisiness, and loudness, and also reverberation, Signal Quality, and Overall quality of each audio clip. In the following, each section of the test template, as seen by participants, is described. These sections are predefined and only the audio clips under the test will be changed from one study to another.

In the first section, the participant’s eligibility and their device suitability are tested and a qualification is assigned to those that pass which remains valid for the entire experiment. The participant's hearing ability is evaluated through digit-triplet-test~\cite{naderi_towards_2020}. Moreover, we test if their listening device supports the required bandwidths (i.e., full-band, wide-band, and narrow-band); details are in Section ~\ref{optimization}).

\begin{figure}
    \centering
  \includegraphics[width = 0.8\columnwidth]{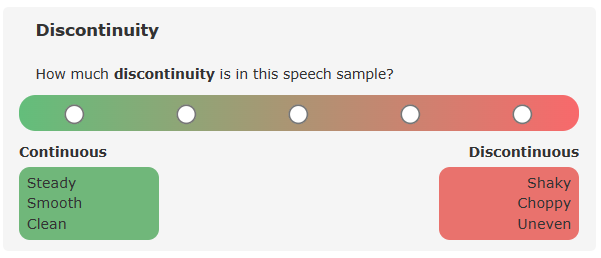}
  \caption{Sub-dimensions are rated on a 5-point discrete scale with descriptive adjectives on poles.}
  \label{fig:scale}
    \vspace{-0.3cm}
\end{figure}

Next, the participant's environment and device are tested using a modified-JND test~\cite{naderi_application_2020} in which they should select which stimulus from a pair has a better quality in four questions. A temporal certificate will be issued for participants after passing this section which expires after two hours and consequently repeating this section will be required. Detailed instructions are given in the next section including introducing the rating scales and providing multiple samples for each perceptual dimension. Participants are required to listen to all samples for the first time. Figure~\ref{fig:scale} illustrates how the rating scale for quality dimensions is presented to participants. In addition, we used a Likert 5-point scale for signal quality and overall quality as specified by ITU-T Rec.~P.835. In the Training section participants should first adjust the playback loudness to a comfortable level by listening to a provided sample and then rate 7 audio clips. This section is similar to the ratings section, but the platform provides live feedback based on their ratings. By completing this section a temporal certificate is assigned to the participants which is valid for one hour. Last is the Ratings section, where participants listen to ten audio clips and two gold standard and trapping questions and cast their votes on each scale. The gold standard questions are the ones that the experimenter already knows their answers (being excellent or bad) and participants are expected to vote on each scale with a minor deviation from known the answer~\cite{naderi_towards_2020}.  Trapping questions are questions in which a synthetic voice is overlaid to a normal clip and asks participants to provide a specific vote to show their attention \cite{naderi_effect_2015}. For this test, we provide scripts for creating the trapping clips, which ask participants to select answers reflecting the best or worst quality in all scales. For rating an audio clip, the participant should first listen to the end of the clip, and then they start casting their votes. During that time, the audio will be played back in a loop. After participants finish with a test set, they can continue with the next one where only the rating section will be shown until other temporal certificates are valid. By the expiration of any certificate, the corresponding section will be shown when they start the next test set.

\begin{table}
    \centering
     \caption{Labels on each scale's pole and descriptive adjectives provided to participants. Terms used in ITU-T Rec.~P.804 are marked in red.}
  \label{tab:terms}
  \includegraphics[width = 1\columnwidth]{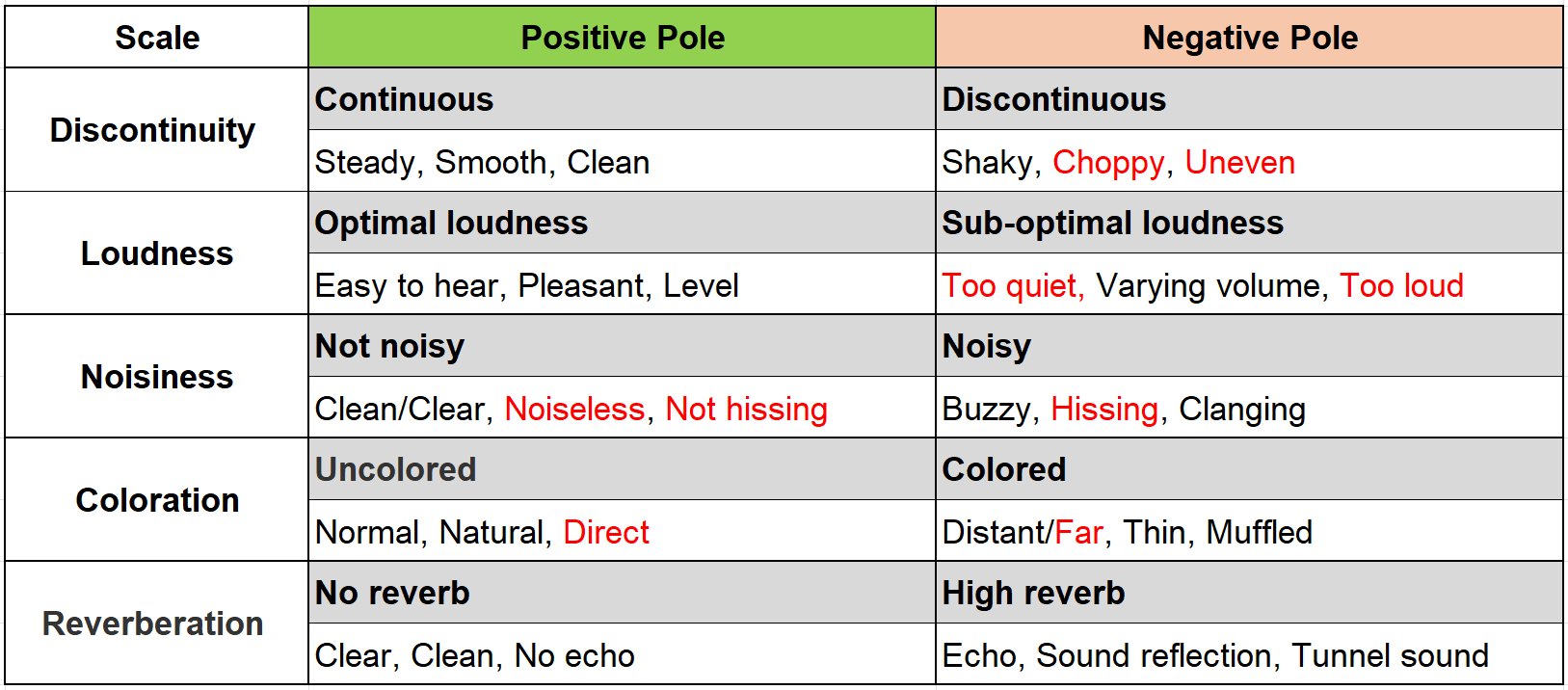}
 
  \vspace{-0.3cm}
\end{table}

%Potential questions from the community:
%\begin{itemize}
%\item Why not use extended scales as in the recommendation?
%\item Why adding SIG and Reverb? What are their utilities? 
%\item What is the difference with the implementation that Gabriel has used?
%\end{itemize}

\vspace{-0.3cm}
\subsection{Survey optimization}
\label{optimization}
We utilized the multi-scale template in various research studies and improved it through the incorporation of experts and test participant feedback.

\textbf{Descriptive adjectives:} The understanding of perceptual dimensions might not be intuitive for naive test participants, therefore the P.804 recommendation includes a set of descriptive adjectives to describe the presence or absence of each quality dimension. We expanded this list through multiple preliminary studies, where participants were asked to listen to samples from each perceptual dimension and name three adjectives that best describe them. For each dimension, we selected the top three most frequently selected terms and presented them below each pole of the scale, as shown in Figure~\ref{fig:scale}. The list of selected terms is reported in Table~\ref{tab:terms}. We used discrete scales for dimensions to be consistent with Signal and Overall scales. 

\textbf{Bandwidth check:} This test ensures the participant devices support the expected bandwidth. The test consists of five samples, and each has two parts separated by a beep tone. The second part is the same as the first part but in three samples superimposed by additive noise. Participants should listen to each sample and select if both parts have the same or different quality. We filtered the white noise with the following bandpass filters: 3.5-22K (all devices should play the noise), 9.5-22k (super-wide-band or fullband is supported), and 15-22K (fullband is supported).

\textbf{Gold questions:}  Gold questions are widely used in crowdsourcing  ~\cite{naderi_towards_2020}. Here we observed gold questions that represent the strong presence of an impairment on one dimension and the clear absence of impairment on another dimension can best reveal an inattentive participant. 
%Table ~\ref{tab:performance_1} shows the effect of filtering submission when filtered based on fails of different gold clip types.

\textbf{Randomization:} We randomize the presentation order of scales for each participant. However, the Signal and Overall quality are always presented at the end. The randomized order is kept for each participant until a new round of training is required.

%\subsection{Extensions to P.804}
\vspace{-0.5cm}
\section{Validation}
\label{sec:validation}
\vspace{-0.2cm}

\subsection{Reproducibility}
\label{reliability}
We used a subset of the blind test set from ICASSP SIG 2023 challenge \cite{cutler_icassp_2023-1} for our reproducibility test. We selected 50 audio clips from the challenge which are processed by 18 models (and the degraded source clips) leading to 950 audio clips. We repeated our crowdsourcing test 5 times with a mutually exclusive group of workers, on separate days on Amazon Mechanical Turk. We calculated the Mean Opinion Scores (MOS) per clip and per model and show the correlation between different runs and scales in clip and model level in Table~\ref{fig:repro}, respectively. The results show a strong correlation between different runs at the model level.

%\begin{figure*}
\begin{table*}[t]
    \centering
    \caption{Pearson correlation between different runs of the reproducibility test in clip and model level.}
  \label{fig:repro}
  \includegraphics[width = 1\textwidth]{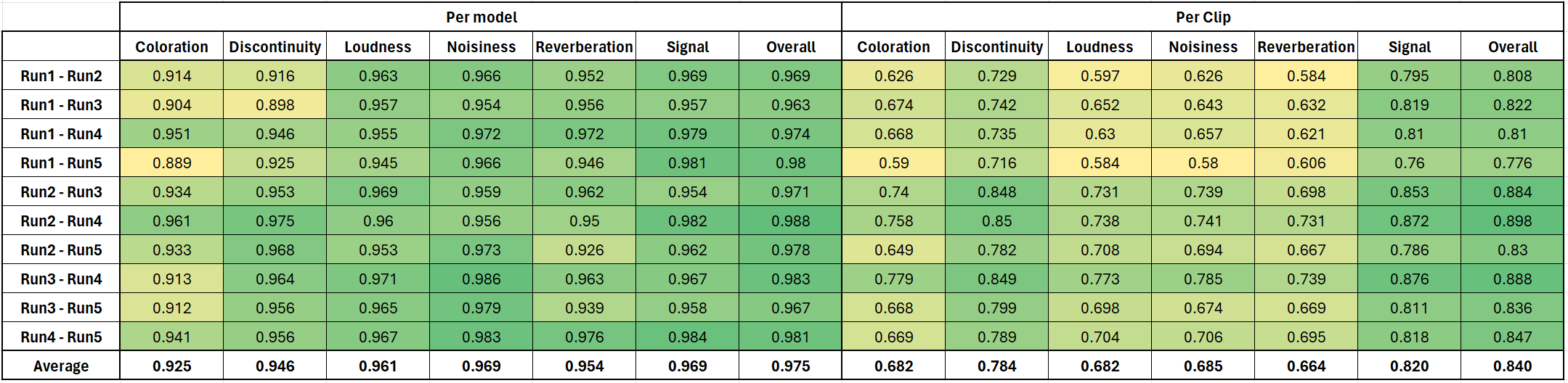}
  
%\end{figure*}
\vspace{-0.5cm}
\end{table*}

\subsection{Accuracy}
In a separate experiment, a subset of data employed in Section~\ref{reliability} is assessed by expert listeners. This subset comprises 20 degraded source clips and their enhanced versions generated by 9 models from the ICASSP SIG 2023 challenge. Table~\ref{tab:expert} presents the correlation between MOS values provided by experts and crowdsourcing, indicating a robust correlation for all dimensions except coloration and reverberation. Further investigation reveals a poor agreement among experts on these dimensions ($ICC_{2k}=.256$, $ICC_{2k}=.267$, for reverberation and coloration, respectively).

\begin{table}
\centering
\caption{Correlation between crowdsourcing P.804 results and expert ratings.} 
\label{tab:expert}

\resizebox{0.8\columnwidth}{!}{%
\begin{tabular}{l c c c c} 
\toprule
Scales &  \multicolumn{2}{c}{Per Model} &  \multicolumn{2}{c}{Per Clip} \\
& PCC& SRCC & PCC & SRCC\\
\midrule
Coloration      & 0.487 & 0.576 & 0.485 & 0.472 \\
Discontinuity   & 0.919 & 0.729 & 0.687 & 0.588 \\
Loudness        & 0.928 & 0.939 & 0.718 & 0.683 \\
Noisiness       & 0.972 & 0.964 & 0.731 & 0.703 \\
Reverberation   & 0.817 & 0.697 & 0.584 & 0.509 \\
Signal          & 0.966 & 0.903 & 0.795 & 0.783 \\
Overall         & 0.991 & 0.842 & 0.842 & 0.844 \\
\bottomrule
\end{tabular}
}
%\vspace{-0.5cm}
\end{table}

%\begin{table}[t]
% values should be updated.
%\caption{Effect of different gold questions on the reliability of collected ratings. Gold questions G1 focus on one direction (all dimensions are MOS 1 or MOS 5), while G2 focuses on multi-dimensions (one dimension is MOS 1, but another one is MOS 5). PCC of Case and Fully Cleaned data in clip level.}
%\label{tab:performance_1} 
%\begin{center}
%\setlength\tabcolsep{1.5pt}
%\vspace{-0.5cm}
%\resizebox{\columnwidth}{!}{%
%    \begin{tabular}{l c c c c c c c c}
    %\begin{tabular}{ccccccccc}
%    \toprule
%    \textbf{Case} & \textbf{\#clips} & \textbf{OVRL} & \textbf{SIG} & \textbf{NOISE} & \textbf{COL} 
%    & \textbf{DISC}  & \textbf{LOUD} & \textbf{REVERB}  
%    \\    
%    \midrule
%    All passed &  \\   
%    No filter & 2927 & 0.90 & 0.90 & 0.92 & 0.91 & 0.80 & 0.93 & 0.92 \\
%    G1 gold passed &  1165 & 0.30 & 0.29 & 0.36 & 0.16 & 0.12 & 0.34 & 0.30 \\
%    G2 gold passed & \\
%    \bottomrule
%    \multicolumn{9}{l}{\#clips: number of clips with more than 3 votes in that group}
 %   \end{tabular}
%}
%\end{center}
%\vspace{-0.5cm}
%\end{table}

%\subsection{Accuracy}
%* Blocked by expert ratings:: note we always expect a deviation between experts and naive participants

%other potentials (I am not so much convinced):
%1) detailed training?
%2)  description of labels!?

\vspace{-0.5cm}
\section{Usage}
\label{sec:usage}
The ICASSP 2023 Speech Signal Improvement Challenge \cite{cutler_icassp_2023-1} aimed to encourage research in enhancing speech signal quality in communication systems, a persistent issue in audio communication and conferencing. Participants were provided with a test and blind sets, and winners were determined through this multi-dimensional subjective test. Both the test and blind sets have 500 samples, encompassing a diverse range of speech distortions, including frequency response distortions, bandwidth limitations, reverberation, and packet loss. Overall 9 teams participated in this challenge and the reported results are based on 9x500 processed clips used in the subjective test.
 
We compared the correlation between quality scores collected using this survey (P.804) and P.835-based \cite{naderi_subjective_2021} subjective tests for all entries, which are reported in Table~\ref{tab:sub-p804-835}. A robust correlation was observed in the shared scores between the two subjective methodologies. Regarding team rankings, the only swap occurred between two teams when utilizing scores from the P.835 test, resulting in a tied rank based on P.804 ratings. 
Moreover, we compute the PCC between the subjective P.804 metrics and the metrics obtained using DNSMOS P.835 \cite{reddy_dnsmos_2022} and NISQA \cite{mittag_nisqa_2021}. The correlations vary from PCC $0.478$ to $0.700$, highlighting the ongoing need for a subjective test to precisely assess speech quality.

In addition, we conducted Explanatory Factor Analysis (EFA) \cite{watkins2018exploratory} to explore the underlying relationships among quality dimensions and assess if there's shared variance among sub-dimensions. We applied the Maximum Likelihood extraction method with Varimax rotation, extracting three factors as suggested by the Scree plot. Bartlett’s test of sphericity yielded a significant result, and the KMO value of 0.65 indicated that the data was suitable for explanatory factor analysis. The factor loadings of quality scores on each factor are shown in Table~\ref{tab:factor_loading}. In total, three factors accounted for $62\%$ of the variance in the data. Factor 1 primarily represented signal quality, with high loadings from Signal, Coloration, and Loudness. Discontinuity formed a separate factor, with some cross-loadings from Signal, suggesting limited shared variance between Discontinuity ratings and both Coloration and Loudness. As anticipated, Noisiness constituted a distinct factor orthogonal to the others, with loading from Reverberation. Considering all mentioned factors, we highlight the importance of adding the signal and reverb dimensions to the P.804, since they contribute to orthogonal factors in a significant percentage.

\begin{table}
\centering
\caption{ Correlations between subjective scores obtained from P.804 and P.835 subjective tests on shared dimensions in model level for all entries. 
Tau-b95 is Kendall Tau-b applied to corrected ranked-order by considering 95\% confidence interval of subjective scores according to \cite{naderi_transformation_2020}.} 
\label{tab:sub-p804-835}
\resizebox{\columnwidth}{!}{%
\begin{tabular}{l c c c c} 
\toprule
Dimension & PCC & SRCC & Kendall Tau-b & Tau-b95 \\
\midrule
Background/Noisiness & 0.964 & 0.926 & 0.825 & 0.853 \\
Signal & 0.954 & 0.933  & 0.801 & 0.914 \\
Overall & 0.965 & 0.940 & 0.825 & 0.822 \\
\bottomrule
\end{tabular}
}
%\vspace{-0.5cm}
\end{table}

%\begin{table}
%\centering
%\caption{The PCC between the subjective P.804 results and the objective metrics estimated %with DNSMOS P.835 \cite{reddy_dnsmos_2022} and NISQA \cite{mittag_nisqa_2021} models.} 
%\label{tab:sub-obj-metrics}
%\setlength\tabcolsep{1.5pt}
%\resizebox{\columnwidth}{!}{%
%\begin{tabular}{l l c c} 
%\toprule
%Subjective metric & Objective metric & \multicolumn{2}{c}{PCC} \\
%& & \small{Clip level} & \small{Model level}\\
%\midrule
%P.804 Overall & DNSMOS P.835 OVRL & 0.695 & 0.884\\
%P.804 Overall & NISQA MOS & 0.681 & 0.766\\
%P.804 Signal & DNSMOS P.835 SIG & 0.656 & 0.799\\
%P.804 Noisiness & DNSMOS P.835 BAK & 0.545 & 0.933\\
%P.804 Noisiness & NISQA NOISE & 0.586 & 0.938\\
%P.804 Coloration & NISQA COLOR & 0.663 & 0.872\\
%P.804 Discontinuity & NISQA DISCONTINUITY & 0.478 & 0.310\\
%P.804 Loudness & NISQA LOUDNESS & 0.700 & 0.784\\
%\bottomrule
%\end{tabular}
%}
%\end{table}

\begin{table}
\centering
\caption{Loading of quality dimensions on three-factor structure using Maximum Likelihood extraction method with Varimax rotation. KMO = 0.65. Factor loading \textgreater0.3 is presented.} 
\label{tab:factor_loading}
\resizebox{0.8\columnwidth}{!}{%
\begin{tabular}{l c c c c} 
\toprule
Quality score & Factor 1 & Factor 2 & Factor 3 \\
\midrule
Coloration & 0.787 & & \\
Discontinuity & & 0.936 & \\
Loudness & 0.476 & & \\
Noisiness & & & 0.742 \\
Reverberation & & & 0.413 \\
Signal & 0.824 & 0.481 & \\

\bottomrule
\end{tabular}
}
\vspace{-0.5cm}
\end{table}

\begin{table}
\centering
\caption{Effects of sub-dimensions on Overall quality, considering Signal quality as mediator.} 
\label{tab:mediation}
\resizebox{0.7\columnwidth}{!}{%
\begin{tabular}{l c c c } 
\toprule
Feature & \multicolumn{3}{c}{Effects} \\
 & Total & Indirect & Direct\\
\midrule
Coloration &    0.427 &  0.352      & 0.102 \\
Discontinuity & 0.343 &  0.277      & 0.066 \\
Loudness &      0.229 &  0.083      & 0.146 \\
Noisiness &     0.146 &  0.047      & 0.099 \\
Reverberation & 0.103 &  0.065      & 0.038 \\
\bottomrule
\end{tabular}
}
\end{table}

Additionally, a majority of the sub-dimensions exert an influence on the quality of the signal and the overall quality. To investigate the indirect and total effects of the sub-dimensions on the overall quality, a mediation analysis was conducted, with signal quality serving as the mediator variable. The outcomes of this analysis are presented in Table~\ref{tab:mediation}, which reveals that Coloration had the highest total effect on the overall quality of this dataset.

\section{Conclusions}
\label{sec:conclusions}
This paper describes an open-source toolkit designed for multi-dimensional subjective speech quality assessment in crowdsourcing. We detail the various sections of the test template and present evidence that the collected ratings obtained using this toolkit are both valid and reproducible. Additionally, it was demonstrated that the toolkit can be used to rank speech enhancement models on large-scale subjective tests, and it can provide insights into the effect of each perceptual dimension on overall quality. 
Results also showed that coloration, discontinuity, and noisiness as three orthogonal factors that other dimensions load onto. Future work includes improving the survey training and scale descriptions to improve coloration and reverberation accuracy with respect to expert raters. 

%Discussion: - audio level normalization (do or not to do!)

%\vfill\pagebreak
\pagebreak

\bibliographystyle{IEEEbib}
\bibliography{IC3-AI, others}
\end{document}